\documentclass{appolb}
\usepackage{graphicx}

\begin{document}
\title{Parton Distribution Functions for discovery physics at the LHC%
\thanks{Presented at XXIX Cracow Epiphany Conference on Physics at the Electron-Ion Collider and Future Facilities}%
}
\author{Amanda Cooper-Sarkar
\address{University of Oxford}
}
\maketitle
\begin{abstract}
At the LHC we are colliding protons, but it is not the protons that are doing the interacting. 
It is their constituents the quarks, antiquarks and gluons-collectively known as partons. 
We need to know how what fractional momentum of the proton each of these partons takes at the energy 
scale of LHC collisions, in order to understand LHC physics. Such parton momentum distributions are known 
as PDFs (Parton Distribution Functions) and are a field of study in their own right. However, it is now 
the case that the uncertainties on PDFs are a major contributor to the background to the discovery of 
physics Beyond the Standard Model (BSM). Firstly in searches at the highest energy scales of a few TeV 
and secondly in precision measurements of Standard Model (SM) parameters such as the mass of the W-boson,
 $m_W$, or the weak mixing angle, $sin^2\theta_W$, which can provide indirect evidence for BSM physics 
in their deviations from SM values.
\end{abstract}
  
\section{Introduction to the determination of PDFs}
PDFs were traditionally determined from meausurements of the differential cross sections of Deep 
Inelastic Scattering. In such processes a lepton is scattered from the proton at high enough energy that 
it sees the parton constituents of the proton. The process is seen as proceeding by the emission of a 
virtual boson from the incoming lepton and this boson striking a quark, or antiquark, within the proton. 
The 4-momentum transfer squared, $q^2$, between the lepton and the proton is always negative 
and the scale 
of the process is given by $Q^2=-q^2$. To calculate the cross sections for these scattering processes we require that $Q^2$ is large enough that we may apply perturbative quantum chromodynamics, QCD. This requires $Q^2 > few~$GeV$^2$.

The formalism is presented here briefly, for a full explanation and references see~\cite{book}. 
The form for the differential cross-section for charged 
lepton-nucleon scattering via neutral current (NC, ie neutral mediating virtual bosons, $\gamma, Z$) is given by
\begin{equation}
\frac {d^2\sigma (l^{\pm}N) } {dxdQ^2} =  \frac {2\pi\alpha^2} {Q^4 x}  
\left[Y_+\,F_2^{lN}(x,Q^2) - y^2 \,F_L^{lN}(x,Q^2)
\mp Y_-\, xF_3^{lN}(x,Q^2) \right],
\label{eq:hq2-NCxsec}
\end{equation}
where $Y_\pm=1\pm(1-y)^2$ and $x$, $y$, $ Q^2$ are measureable kinematic variables given in terms of the 
scattered lepton energy and scattering angle and the incoming beam momenta of lepton and 
proton. The three structure functions, $F_2, F_L, xF_3$, depend on the nucleon structure as 
follows, to leading order (LO) in perturbative QCD. (Here by leading order we mean to zeroth order in $\alpha_s(Q^2)$)
\begin{equation}
 F_2^{lN}(x,Q^2) = \sum_i A_i^0(Q^2)*(xq_i(x,Q^2) + x\bar q_i (x,Q^2)),
\label{eq:hq2-f2ln}
\end{equation}
where, for unpolarised lepton scattering,
\begin{equation}
        A_i^0(Q^2) =  e_i^2-2e_iv_iv_eP_Z+(v_e^2+a_e^2)(v_i^2+a_i^2)P_Z^2 
\label{eq:unpol-a}
\end{equation}
\begin{equation}
F_L^{lN}(x,Q^2) = 0,
\end{equation}
and
\begin{equation}
  xF_3^{lN}(x,Q^2) = \sum_i B_i^0(Q^2)*(xq_i(x,Q^2) - x\bar q_i (x,Q^2)),
\label{eq:xf3ln}
\end{equation}
where, 
\begin{equation}
        B_i^0(Q^2) =  -2e_ia_ia_eP_Z+4a_iv_iv_ea_eP_Z^2
\label{eq:unpol-b}
\end{equation}
The term in $P_Z$ arises from $\gamma Z^0$ interference and the term in
$P_Z^2$ arises purely from $Z^0$ exchange, where $P_Z$ accounts for the effect 
of the $Z^0$ propagator relative to that of the virtual photon, and is given by
\begin{equation}
           P_Z = \frac {Q^2} {Q^2 + M_Z^2} \frac {1} {\sin^2 2\theta_W}.
\end{equation}
The other factors in the expressions for $A_i^0$ and $B_i^0$ are
the quark charge, $e_i$, NC electroweak vector, $v_i$, and axial-vector,$a_i$, 
couplings of quark $i$ and the corresponding NC electroweak 
couplings of the electron, $v_e,a_e$. 
At low $Q^2 ( << M_W^2)$ only the $A_i$ term is sizeable and it depends only on 
the quark-charge-squared, see Eq~\ref{eq:hq2-f2ln}.
In the simple Quark Parton Model the structure functions depend ONLY on the kinematic variable $x$, 
so the structure functions scale, and $x$ can be 
identified as the fraction of the proton's momentum taken by the struck quark or antiquark. QCD improves
 on this prediction by taking into account the interactions of the partons, such that a quark can radiate a
 gluon before, or after, being struck and indeed a gluon may split into a quark-antiquark pair and one 
of these is the struck parton. This modification leads to the dependence of the structure functions on 
the scale of the probe, $Q^2$, as well as on $x$. However, this dependence, or scaling deviation, 
is slow, it is logarithmic and is calculated through the DGLAP evolution equations.
We can already see from the equations that measuring the structure functions will give us information 
on quarks and antiquarks, but measuring their scaling violations will also give us information on the 
gluon distribution and furthermore, if we calculate beyond leading order 
we will also see that the longitudinal 
structure function depends on the gluon distribution. 
If we also consider charged current (CC) lepton scattering via the $W^+$ and $W^-$ bosons we find that we 
tell apart u- and d-type flavoured quarks and antiquarks. Scattering with neutrinos rather than charge lepton 
probes give similar information.

The current state of the art is calculations to NNLO in QCD. At this order the relation of the structure 
functions to the 
parton distributions becomes a lot more complicated. However it is completely caluclable so that, given 
the parton distributions at some low scale $Q^2_0$, we can evolve them to any higher scale using the NNLO 
DGLAP equations and then calculate the measurable structure functions using the NNLO relationships 
between the structure functions and the parton distributions.  This allows us to confront these 
predictions with the measurements. But how do we know the parton distribution functions  
at $Q^2_0$. We don't! We have to parametrise them. The parameters are then fitted in a $\chi^2$ fit of 
the predictions to the data. Given that there are typically $\sim5000$ data points and $\sim25$ parameters, the 
success of such fits is what has given us confidence that QCD IS the theory of the strong interaction.
 
\section{Uncertainties on PDFs and consequences for the LHC}
Several groups worldwide perform these sort of fits to determine PDFs. In doing so they make different 
choices about paramterisations, about model inputs to the calculation and about methodology. 
PDFs are typically parametrised at the starting scale by
\begin{equation}
xf_i(x) = A_i x^{B_i} (1-x)^{C_i} P_i(x), f_i= {u,\bar{u},d,\bar{d},s,\bar{s},g}
\end{equation}
where $P_i(x)$ is a polynomial in $x$ or $\sqrt{x}$, which could be an ordinary polynomial, a Chebyshev 
or Bernstein polynomial, or indeed $P_i(x)$ could actually be given by a neural net.
Some parameters are fixed by the number and momentum sum-rules, but for others chosing to fix or free 
them constitute model choices. For example; the heavier quarks are often chosen to be generated by gluon splitting, 
but they could be parametrised; the strange and 
antistrange quarks can be set equal, or parametrised separately. 
Other choices are the value of the starting scale $Q^2_0$; the choice of data accepted for the fit and 
the kinematic cuts applied to them; the choice of heavy quark scheme and the choice of heavy 
quark masses input. Although the HERA collider DIS data~\cite{HERAPDF20} form the backbone of all modern PDF fits, 
earlier DIS fixed-target data has also been used as well as Drell-Yan data from fixed target scattering and 
in particlar, $W$ and $Z$ production data from the Tevatron and indeed from the LHC. 
High $E_T$ jet data from both Tevatron and LHC have also been used and more recently LHC $t\bar{t}$ production 
data, $Z p_T$ spectra, $W$ or $Z$ + jet data and $W$ + heavy flavour data have all been used.

\begin{figure}[htb]
\includegraphics[width=6.5cm]{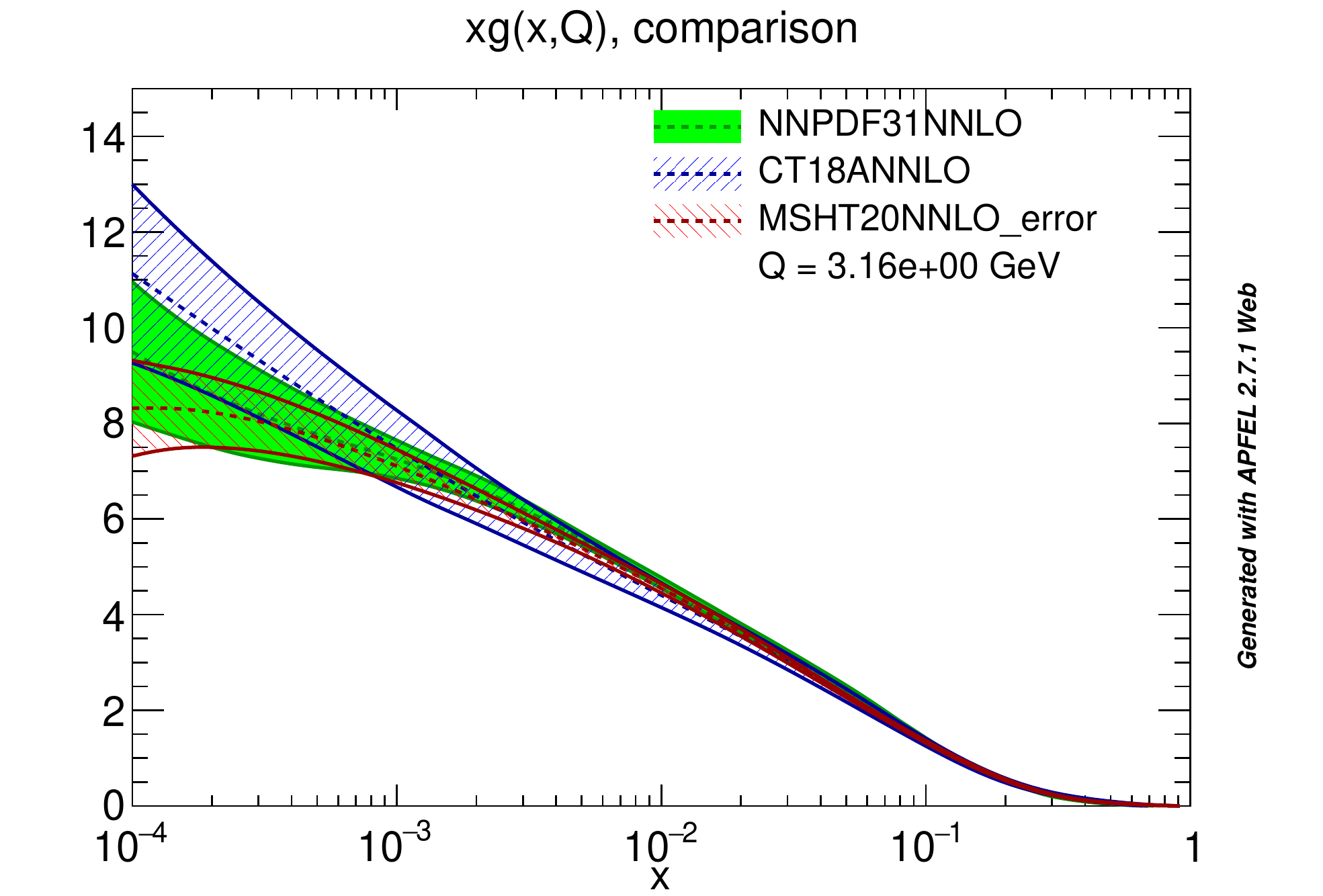}
\includegraphics[width=6.5cm]{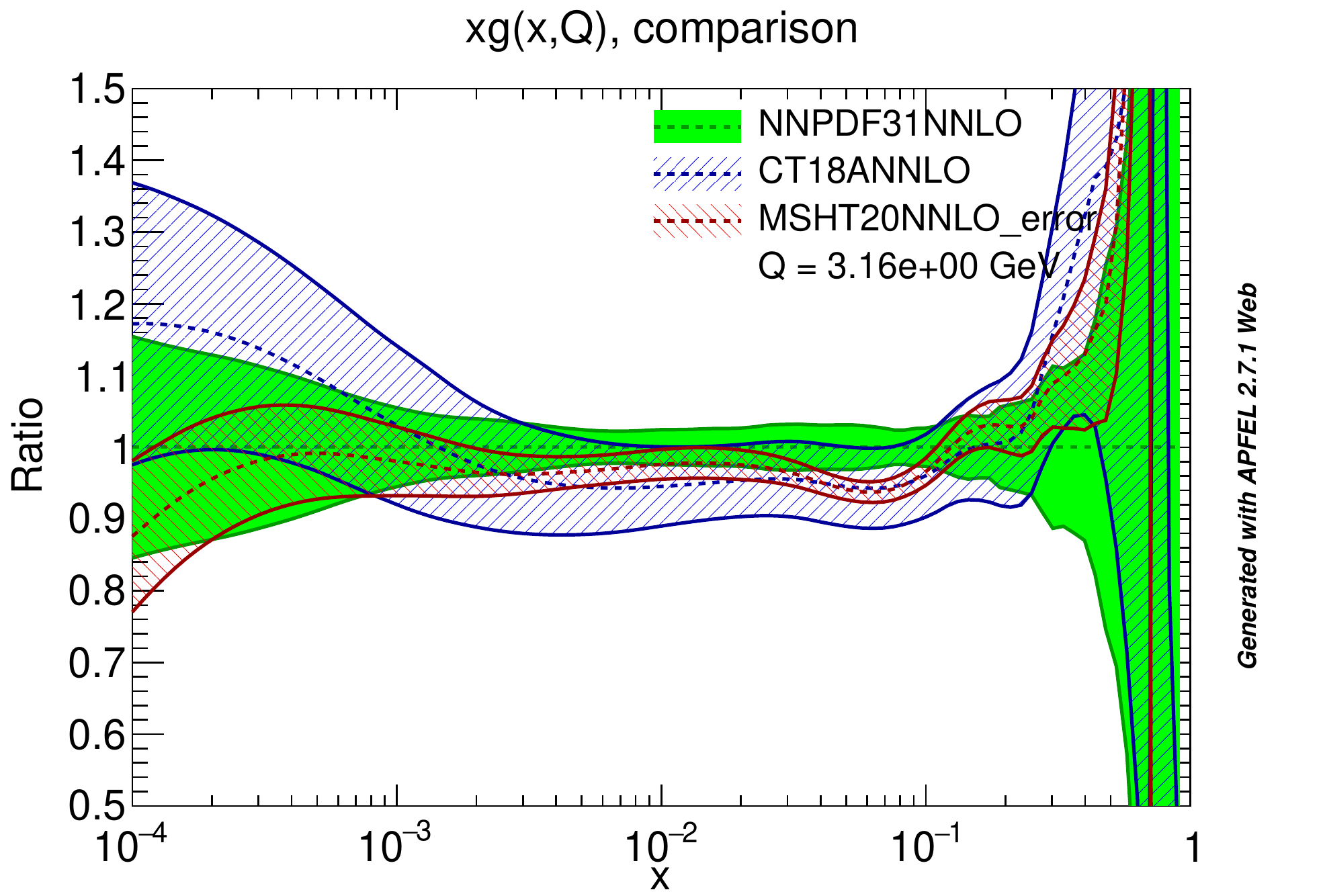}
\includegraphics[width=6.5cm]{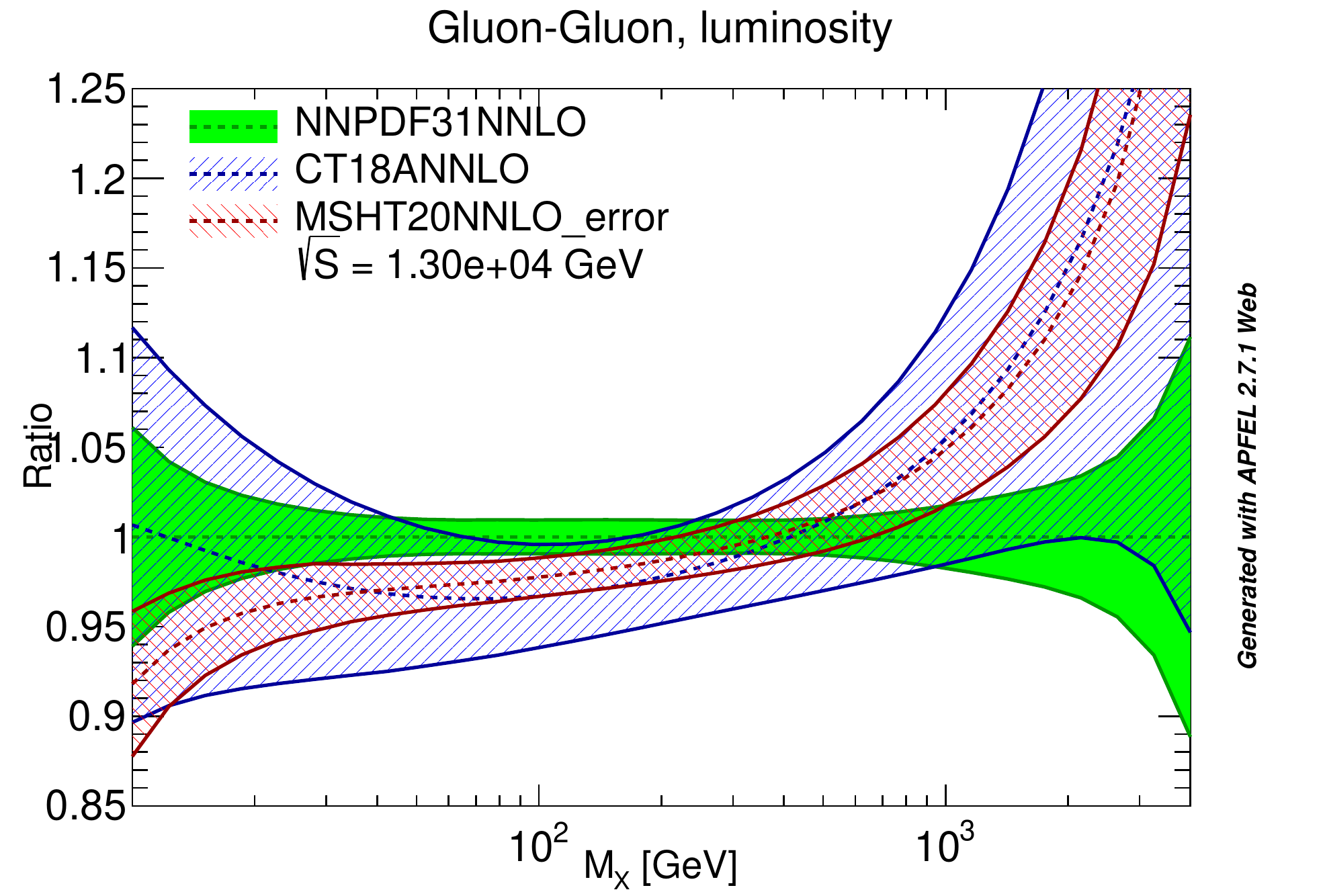}
\includegraphics[width=6.5cm]{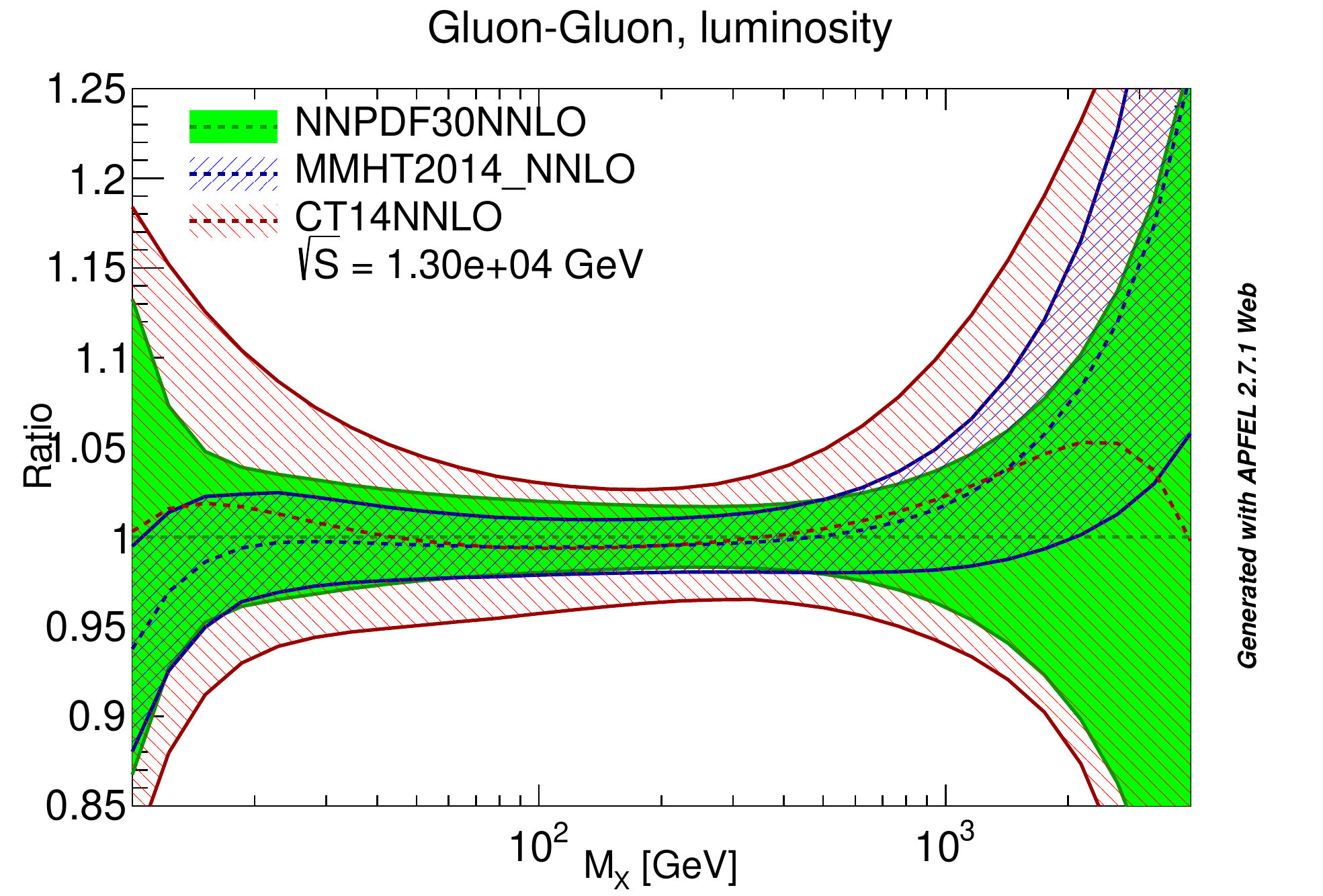}
\caption{Gluon distributionss of NNPDF3.1, MSHT20, CT18 compared (top left), compared in ratio to NNPDF31 
(top right). Gluon-gluon luminosities compared for NNPDF3.1, MSHT20, CT18 (bottom left) and NNPDF3.0, MMHT14, CT14 (bottom riht)}
\label{fig:pdfs}
\end{figure}

Given that groups make different choices, how are we doing? Fig~\ref{fig:pdfs} (top left) show comparisons of the 
latest NNLO gluon PDFs from the three global PDF fitting groups, 
NNPDF3.1, CT18A, MSHT20~\cite{3pdfs} at a typical low scale \footnote{Other notable PDF analyses are HERAPDF2.0~\cite{HERAPDF20}, ABMP16~\cite{abmp16}, and ATLASpdf21~\cite{2112.11266} but none of these include such a wide range of data}. Looking at this plot we have the impression that the 
level of agreement between the three groups is not bad. However, if we look at the ratio of the gluon pdfs to that of NNPDF3.1 in 
Fig~\ref{fig:pdfs} (top right), we see that the situation is only good (within $\sim 5\%$) at middling $x$. Disagreement at 
low and high $x$ is quite signficant. Although this is illustrated only for the gluon, the situation is similar for 
all PDFs.

To see how this affects physics at the LHC we must first consider how these cross sections are calculated in order 
that we can make sense of a definition of parton-parton luminosities.
\begin{equation}
d\sigma_{hard}(p_A,p_B,Q^2) =\sum_{ab}\int dx_a dx_b 
f_{a/A}(x_a,\mu^2)f_{b/B}(x_b,\mu^2)\nonumber\\
. d\sigma_{ab\to cd}(\alpha_s(\mu^2),Q^2/\mu^2)
\label{eqn:had-had-master}
\end{equation}
where $d\sigma_{ab\to cd}$ is the parton-parton cross-section at a hard 
scale $Q^2$ and $f_{a/A}$ is the parton momentum density of parton $a$ 
in hadron $A$ at a factorisation scale $\mu^2$ (and similarly for $b,B$). The initial parton momenta 
are $p_a=x_ap_A,~p_b=x_bp_B$. The hard
scale $Q^2$ could be provided by jet $E_T$ or Drell-Yan lepton-pair mass, for example.
Strictly the 
scale involved in the definition of $\alpha_s$ in the cross-section (the
renormalisation scale) could be different from the factorisation scale for the partons, but
it is usual to set the two to be equal and indeed the choice $\mu^2=Q^2$
is often made. We have assumed the factorisation theorem.
A parton-parton luminosity is the normalised convolution of just the parton distribution part of the above 
equation for LHC cross-sections~\cite{ppluminositydef}. The gluon-gluon luminosities for NNPDF3.1, MSHT20 and CT18 
is shown in the bottom left part of Fig~\ref{fig:pdfs} in ratio to NNPDF3.1. The $x$-axis is the c.of.m energy of the 
system $X$ which is produced in the gluon-gluon collision, $M_X = \sqrt(x_a x_b s)$.  
We can see that the luminosities are in good agreement at the the Higgs mass, but less so at smaller 
and larger scales. 

\begin{figure}[htb]
\includegraphics[width=6.5cm]{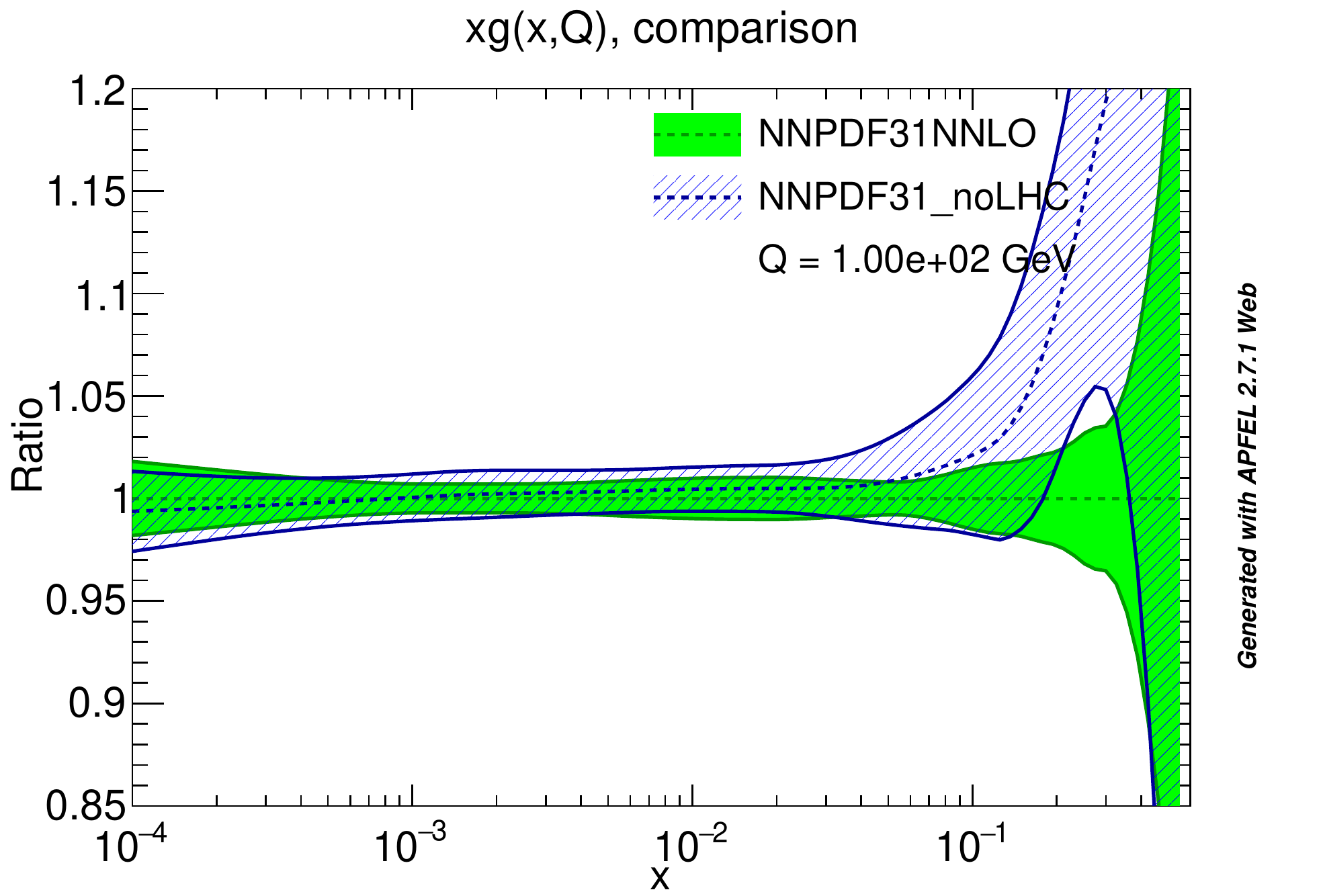}
\includegraphics[width=6.5cm]{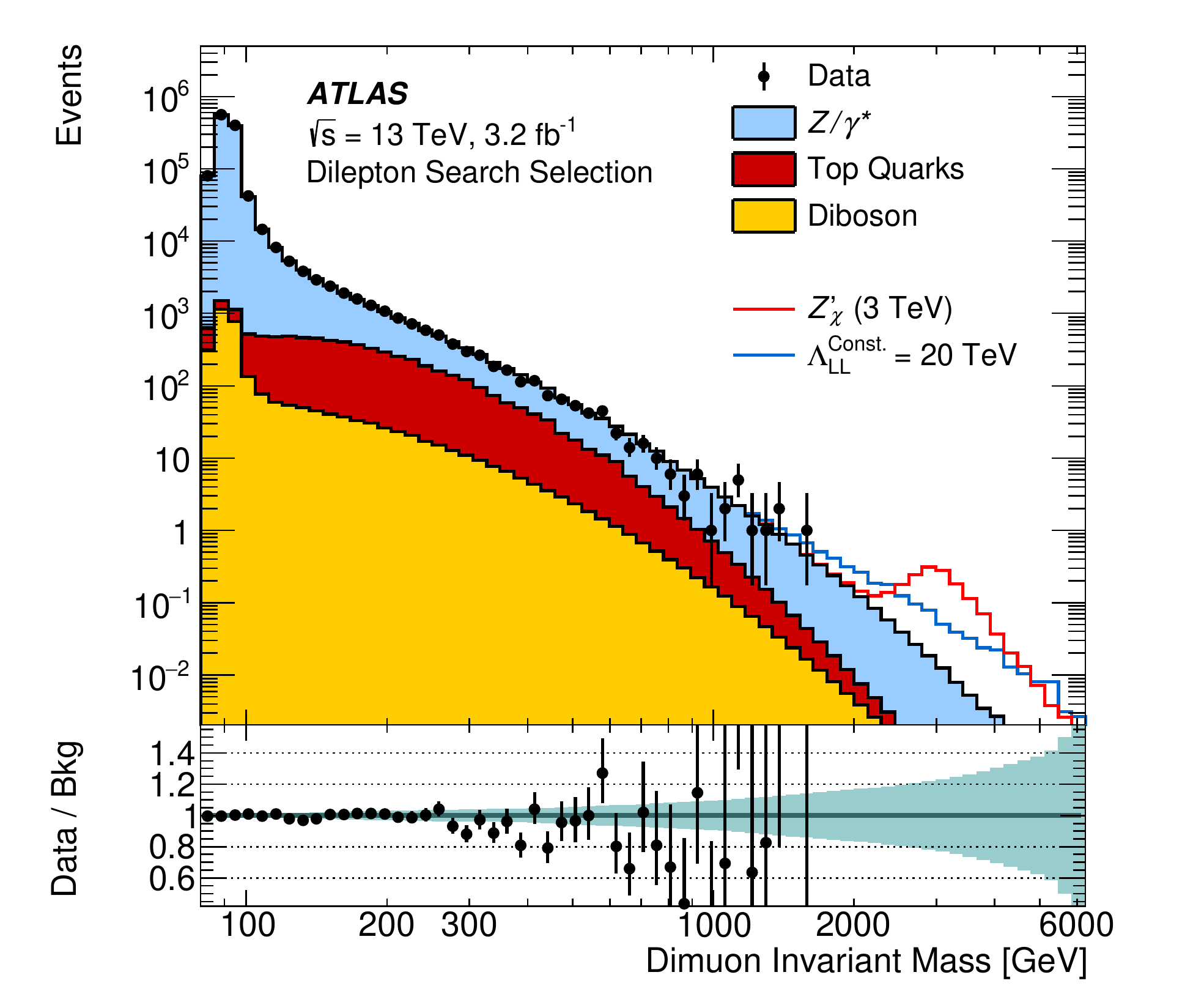}
\caption{Left; Gluon distributions for NNPDF3.1 with and without LHC data: Right; ATLAS dilepton mass spectrum from 13 TeV data~\cite{1607.03669}, illustrating the effects of a 3 TeV $Z'$ 
or a 20 TeV contact interaction. The ratio panel beneath shows data over background with ths systematic 
uncertainty of the measurement shown in grey}
\label{fig:bsm}
\end{figure}

We may ask the question has the LHC data decreased the uncertainty on the PDFs. In Fig.~\ref{fig:bsm} (left) we 
compare the NNPDF31 gluon distribution with and without the LHC data in ratio. We can see that the LHC data has decreased 
the uncertainty and changed the shape. But we cannot draw a conclusion on the basis of one PDF alone. The NNPDF3.1 
analysis makes specific choices of which jet production data to use, which $t\bar{t}$ distributions to use etc., 
and specific choices of how to treat the correlated systematic uncertainties for these data. Other PDF fitting 
groups make different choices. We need to look at the progress made by all three groups. Fig.~\ref{fig:pdfs} 
(bottom right) shows a comparison of the gluon-gluon luminosity for all three groups for the previous generation of PDFs, 
NNPDF3.0, MMHT14, CT14~\cite{3pdfsold}, for which very little LHC data were used. If we compare this with the recent gluon-gluon 
luminosity plot in Fig.~\ref{fig:pdfs} (bottom left) we see that, whereas each group has reduced its uncertainties, their 
central values were in better agreement at the Higgs mass for the previous versions!
Thus analysis of new data can introduce discrepancies.

An effort to combine the three PDFs, called PDF4LHC15,  was performed for the previous versions, and a new 
combination, called PDF4LHC21~\cite{pdf4lhc21}, as been performed for the most recent versions. The combination 
procedure uses MC replicas from all three PDFs and then compresses them with minimal loss of 
information. Altough the overall uncertainties of PDF4LHC21 are smaller than those of PDF4LHC15. 
The improvement is not as dramatic as one might 
have hoped, precisely because of deviation in central values.
Since PDF4LHC21 the NNPDF group have updated to NNPDF4.0~\cite{nnpdf40}, which has considerably reduced uncertainties compared 
to NNPDF3.1. However, this is mostly due to a new methodology rather than due to new data. Unfortunately this puts 
the NNPDF4.0 central values further from those of CT and MSHT in some regions, such that there is no big 
improvement in the combination of all three.


To illustrate the impact for direct searches for BSM physics from PDF uncertainties Fig.~\ref{fig:bsm} (right) illustrates 
two types of searches done in dilepton production using 13 TeV ATLAS data, one for a 3 TeV $Z'$ and one for 
contact interactions at 20 TeV. The panel below the main plot shows the ratio of data to SM background and the 
grey uncertainty and on this shows the projected systematic uncertainty band of the measurement. A major 
contributor to this uncertainty is the PDF uncertainty of the background calculation. Whereas a resonant $Z'$ at 
higher scale could likely be distinguished from the background this is far less clear for the gradual change in 
shape induced by the contact interaction. Indeed this could potentially be accommodated in small changes to the 
input PDF parameters such that it would remain hidden.

Given that no searches for BSM physics at high scale have given a significant signal, the effects of BSM are also 
investigated indirectly by making precision measurements of SM parameters such as the mass of the W-boson,
 $m_W$, or the weak mixing angle, $sin^2\theta_W$, which can provide indirect evidence for BSM physics 
in their deviations from SM values. For example, $m_W$ is predicted in terms of other SM parameters but there is a 
contribution from higher order loop diagrams which would include any BSM effects. This would raise the value of 
$m_W$ noticeably if the scale of the BSM effects is not very far above presently excluded limits. The recent 
meausurement from CDF~\cite{cdfmw} is $m_W = 80.433 \pm 0.009$ GeV, well above the SM prediction of 
$80.357 \pm 0.006$ GeV. However, 
many other measurements are not discrepant and the next most accurate is the ATLAS 7 TeV measurement 
$m_W = 80.370 \pm 0.019$ GeV. 
Obviously, one would like to improve the accuracy on this LHC measurement, but a major 
part of the 19 MeV uncertainty is $\sim 10$ MeV coming from PDF uncertainty. The LHC PDF uncertainty is larger 
than that of CDF because the LHC 7 TeV $pp$ collisions are mostly sea quark-antiquark collisions at $x\sim 0.01$, 
whereas the CDF $p\bar{p}$ collisions are mostly valence-valence collisions at $x\sim 0.07$. 
To substantially reduce the LHC PDF uncertainty one requires PDF uncertainties of $O(1\%)$ 
in the relevant $x$ range.

The vital question for both direct and indirect searches is whether the PDF uncertainty can be reduced in future. 
A study of potential improvements from the High-Luminosity Phase of the LHC was made~\cite{hllhc} asssuming a 
luminosity of $3 ab^{-1}$ of data. The processes considered were those which have not yet reached the limit in 
which the data uncertainties are systematic dominated e.g. 
higher mass Drell-Yan, $W+c$, direct photon, $Zp_T$ at very 
high $p_T$, higher scale jet production and higher scale $t\bar{t}$ production. Two different sets of assumptions 
were made about the systematic uncertainties, pessimistic and optimistic. For the pessimistic case there is no improvement in systematic uncertainties, for the
optimistic case one assumes that a better knowledge of the data gained from higher statistics could result in a 
reduction of the size of systematic uncertainties by a factor of 0.4, and in the role of correlations between 
systematics uncerrainties by a 
factor of 0.25. Improvements of in the PDF uncertainties of about a factor of two are predicted for gluon, 
$u$ and $d$ quarks and antiquarks. Whereas this looks very 
promising -reducing the PDF4LHC PDF uncertainty from $\sim 4\%$ to $\sim 2\%$ over the relevant $x$ range for $m_W$ 
measurement, with little difference in pessimistic and optimistic assumptions- we should 
remember that a) we aim for $O(1\%)$ 
accuracy on PDFs, and b) such a pseudo-data analysis is necessarily over-optimistic in that it assumes that the 
future data are fully consistent with each other and that systematic uncertainties have well-behaved Gaussian 
behaviour. In reality this is never the case - this is why $\Delta\chi^2$ tolerance, $T$, values in the CT and MSHT 
analyses are set at $T >\sim 3$.

\begin{figure}[htb]
\includegraphics[width=6.5cm]{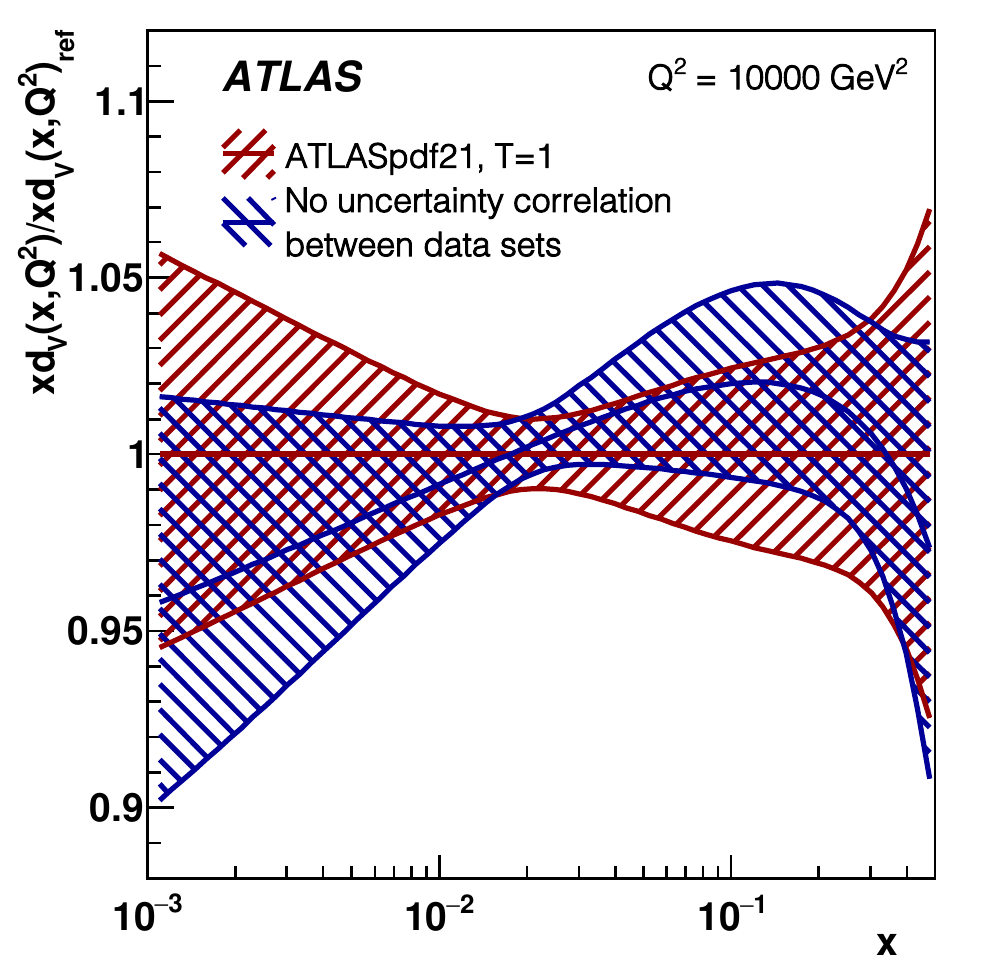}
\includegraphics[width=6.5cm]{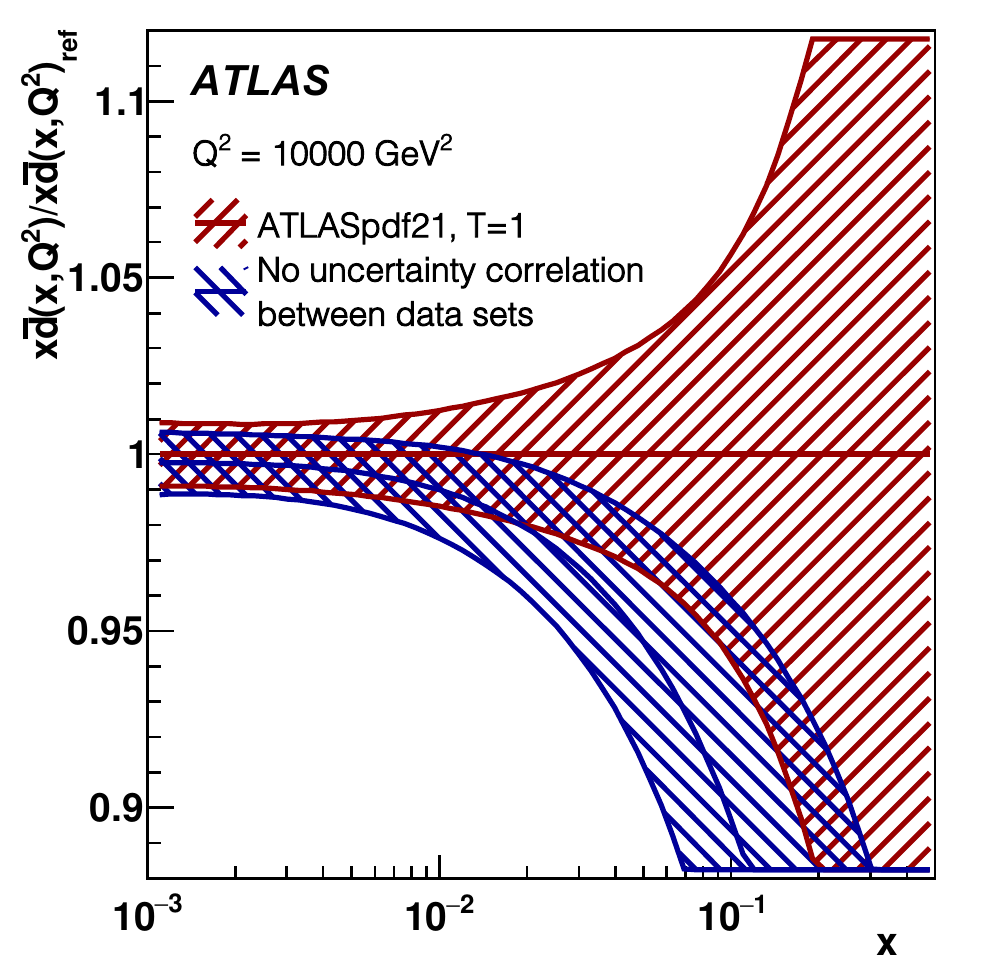}
\includegraphics[width=6.5cm]{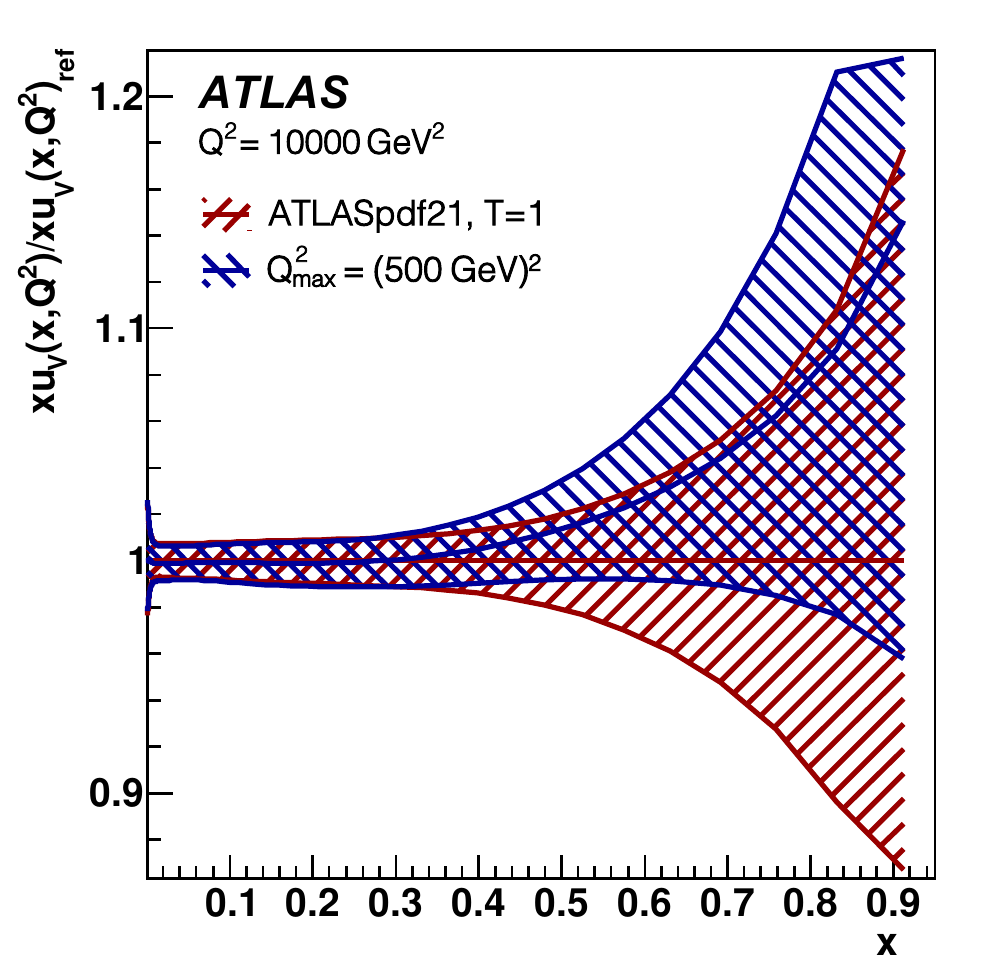}
\includegraphics[width=6.5cm]{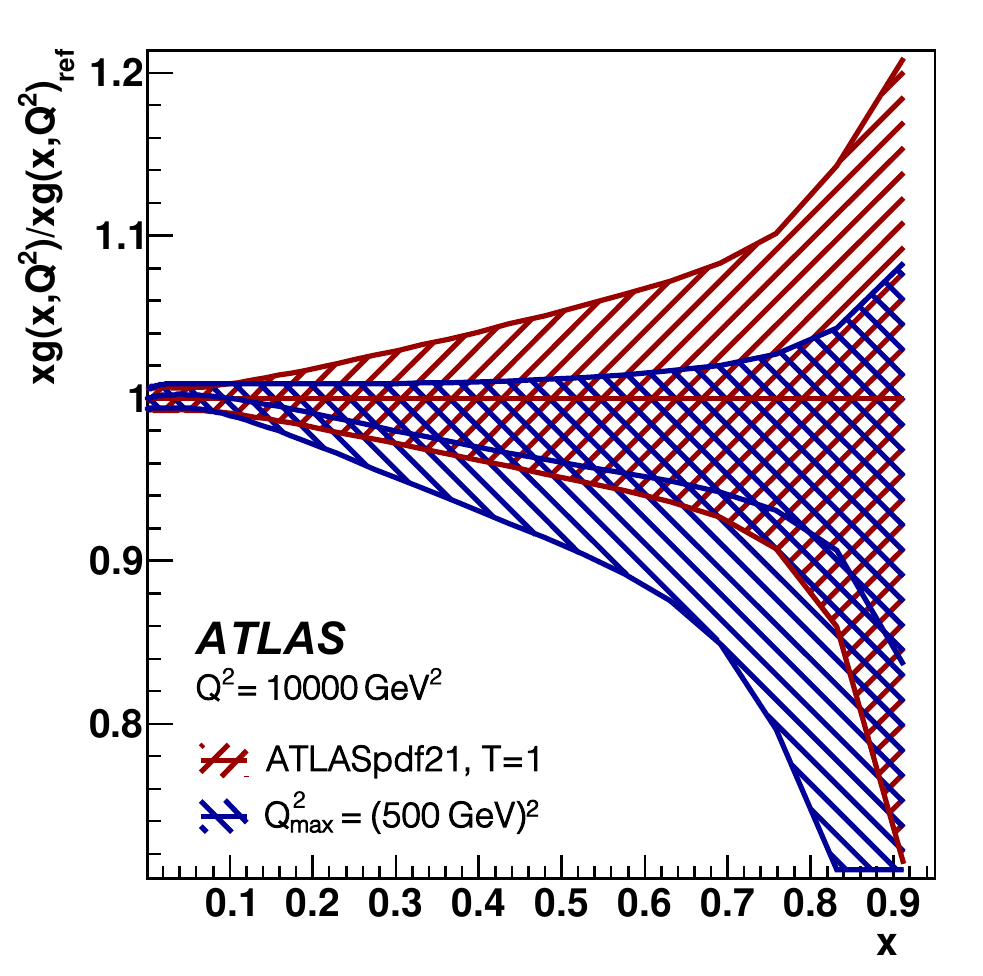}
\caption{ATLASpdf21 $xd_v$ (top left) and $x\bar{d}$ (top right) PDFs accounting for interdata set systematic correlations in ratio to 
those obtained not accounting for these correlations. ATLASpdf21 $xu_v$ (bottom left) and $xg$ (bottom right) PDFs cutting high scale data 
$Q>500$GeV in ratio to not cutting these data. Note the linear $x$ scale emphasizes high-$x$}
\label{fig:ATLAS}
\end{figure}

A further issue highlighted by a recent ATLAS PDF analysis~\cite{2112.11266} is that there can be correlations of 
systematic uncertainties between data sets as well as within them. The ATLAS analysis used many different types of 
ATLAS data. Amongst these were inclusive jets, $W$ and $Z$ boson + jets, $t\bar{t}$ in lepton + jet mode. 
The systematic uncertainties on the jet measurements are correlated between these data sets and an egregious example 
of this is the relatively large uncertainties on the jet energy scale. The ATLAS analysis showed that the 
difference in the resulting PDFs between accounting for these correlations and not accounting for them can exceed 
$1\%$ at the relevant energy scale and $x$ region for $W$ production, see Fig~\ref{fig:ATLAS} (top part). 
Thus PDFs cannot become $1\%$ accurate without
 accounting for such correlations. The information needed to do this was not avalable to the global PDF 
fitting groups prior to this ATLAS analysis, and it needs to become available for many more data sets included 
in their fits.

This ATLAS analysis also made a study in which data at very high scale $Q>500$ GeV ($Q^2 >250000$ GeV$^2$) are cut. Most of the data cut are the high-$p_T$ jet production data. If new physics at high 
scale makes a subtle change to the shape of jet high-$p_T$ spectra, it will also make a change to the PDF parameters when fitted. 
Thus there may 
be a difference in PDFs fitting or not fitting high scale data. Fig~\ref{fig:ATLAS} (bottom part) shows such a comparison for the gluon and $xu_V$ 
PDFs, which are most affected by this cut. There is no significant difference even at very high-$x$.

A further limitation on PDF accuracy is scale uncertainty. The ATLASPDF21 analysis included scale uncertainties on 
the NNLO predictions for inclusive $W,Z$ production, which is the only process included for which these 
uncertainties are comparable to the experimental unceratinties, for the other processes scale uncertainties are 
significantly smaller. Comparing the PDFs with and without accounting for these scale uncertainties showed that 
$\sim 1\%$ discrepancies in central PDF values can also come from this source. But the situation may be worse 
than this. MSHT have recently performed an approximate N3LO analysis~\cite{MSHTaN3LO}. The MSHT20 N3LO and NNLO gluon differences 
are very strong at low-$x < \sim 10^-3$ and low scale and this difference persists to LHC scales such that there 
is still a 
$\sim 5\%$ discrepancy at $Q^2=10000$GeV$^2$ and $x \sim 0.01$. This translates into a $5\%$ difference in 
gluon-gluon luminosity at the Higgs mass! This difference is a consequence of the much stronger differences at 
low-$x$ and low $Q^2$ and such differences will matter more as we go to higher energies and/or to more forward physics at the 
LHC. However, we will also need an improved theoretical understanding of low-$x$  physics, such as $ln(1/x)$ resummation and non-linear 
effects due to parton recombination, to fully exploit this region, see for example~\cite{lowx} and references therein.

So how could we improve the PDFs in future? A dedicated lepton-hadron collider would provide the most accurate PDFs. The reason that a lepton-hadron collider can improve PDF uncertainty more than a hadron-hadron machine is that 
the inclusive DIS process, from which most of the information comes, can be analysed by a single team, with a consistent 
treatment of systematic 
uncertainties across the whole kinematic plane. This situation does not appertain at the LHC where different teams 
analyse the many different processes which are input to the PDF fits. Whereas there are common conventions for measurement, complete consistency is 
rarely obtained, particularly since the optimal treatment of data evolves with time and analyses proceed at different paces.

Proposals for an LHeC or even an FCC-eh machine at CERN have been made and these would improve the PDFs very 
substantially across a kinematic region ranging down to $x \sim 10^{-6} (10^{-7})$ and up to 
$Q^2 \sim 10^6 (10^7)$ for the LHeC (FCCeh) respectively~\cite{klein}. Such a collider will also be able to shed light on low-£x£ physics.
The EIC collider at Brookhaven is an approved project which will extend the kinematic region of accurate 
measurement to higher $x$ at low scales~\cite{hobbs} and this would benefit studies at the LHC scales, firstly because DGLAP 
evolution percolates from high to low $x$ as the scale increases and secondly because the momentum sum-rule ties all $x$ regions together.

\section{Summary}
The precision of present PDFs needs improvement in order to aid discovery physics, both at high scale and in the precision measurement of 
SM parameters. Substantial improvement should come from the HL-LHC run, but the desired accuracy of $O(1\%)$ can only be achieved at a 
future lepton-hadron collider. The EIC with improve accuracy at high $x$, but for low-$x$ physics an LHeC or FCC-eh is necessary.

\end{document}